\documentclass[conference]{IEEEtran}
\IEEEoverridecommandlockouts
\usepackage{cite}
\usepackage{amsmath,amssymb,amsfonts}
\usepackage{algorithmic}
\usepackage{graphicx}
\usepackage{textcomp}
\usepackage{xcolor}
\def\BibTeX{{\rm B\kern-.05em{\sc i\kern-.025em b}\kern-.08em
    T\kern-.1667em\lower.7ex\hbox{E}\kern-.125emX}}
\begin{document}

\title{Integrating Preprocessing Methods and Convolutional Neural Networks for Effective Tumor Detection in Medical Imaging\\

}

\author{\IEEEauthorblockN{Ha Anh Vu}
\textit{Central Washington University }\\
HaAnh.Vu@cwu.edu}

\maketitle

\begin{abstract}
This research presents a machine-learning approach for tumor detection in medical images using convolutional neural networks (CNNs). The study focuses on preprocessing techniques to enhance image features relevant to tumor detection, followed by developing and training a CNN model for accurate classification. Various image processing techniques, including Gaussian smoothing, bilateral filtering, and K-means clustering, are employed to preprocess the input images and highlight tumor regions. The CNN model is trained and evaluated on a dataset of medical images, with augmentation and data generators utilized to enhance model generalization. Experimental results demonstrate the effectiveness of the proposed approach in accurately detecting tumors in medical images, paving the way for improved diagnostic tools in healthcare.
\end{abstract}


\section{Introduction}
Tumor detection in medical images is crucial in early diagnosis and treatment planning for various diseases. Traditional methods often rely on manual inspection by medical professionals, which can be time-consuming and prone to human error. In recent years, machine learning techniques and intense learning approaches using convolutional neural networks (CNNs) have shown promising results in automating tumor detection tasks. In this study, we aim to develop a machine-learning model for tumor detection using CNNs, focusing on preprocessing techniques and model training.

The research begins with thoroughly exploring preprocessing techniques to enhance image features relevant to tumor detection. These techniques include Gaussian smoothing, bilateral filtering, and K-means clustering, which are applied to preprocess the input medical images and highlight tumor regions. Subsequently, a CNN model is developed using the Keras library, comprising convolutional layers, max-pooling layers, and fully connected layers. 

\section{Literature Review}

Medical imaging, particularly in brain tumor detection, has experienced a significant surge in research that leverages technological advancements to enhance diagnostic accuracy. A pivotal element across these studies is the use of the Figshare dataset\cite{figshare}, demonstrating its critical role in developing innovative diagnostic methodologies. This review aims to synthesize the collective advancements, emphasizing the transition from traditional diagnostic methods to sophisticated computational models.

\begin{itemize}
    \item Khan et al.\cite{brain1} tackle the complexity of brain tumor classification through deep learning, developing CNNs for binary and multiclass categorization. Their methodology exemplifies the potential of deep learning to streamline diagnostics by bypassing manual feature extraction, thus significantly advancing the automation of tumor detection.
    
    \item Soumik and Hossain\cite{brain2} further the field by employing transfer learning with the InceptionV3 model, adeptly handling the scarcity of brain MRI images. Their method's notable accuracy underlines transfer learning's effectiveness in optimizing deep learning models amidst dataset constraints.
    
    \item Irmak\cite{brain3} introduces a fully optimized CNN framework for multi-classification, achieving unparalleled accuracy. Beyond demonstrating CNNs' efficacy, this research pioneers in the algorithmic fine-tuning of hyper-parameters, thereby refining the diagnostic tools available to the medical community.
    
    \item The multi-kernel SVM approach by Dheepak, Christaline, and Vaishali\cite{brain4} innovates brain tumor classification by integrating diverse kernel functions, enhancing the model's robustness. This methodological innovation opens new pathways for improving diagnostic accuracy beyond traditional SVM models.
    
    \item Montoya, Rojas, and Vásquez\cite{brain5} critically assess the performance of shallow versus deep neural networks, with the ResNet50 architecture showcasing superior classification capabilities. Their work corroborates the pivotal role of deep learning in the future of medical diagnostic research.
\end{itemize}

\section{Methodology}
\subsection{Dataset Description}
This study employs the Figshare \cite{figshare} brain MRI image dataset, characterized by its imbalanced distribution among three tumor types: Meningioma (708 slices), Glioma (1426 slices), and Pituitary tumor (930 slices), culminating in 3064 T1-weighted contrast-enhanced images from 233 patients. To facilitate a thorough analysis and classification of brain tumors, the dataset was divided, allocating 70\% of the images for training and the remaining 30\% for testing purposes. Each image is standardized to a 512x512 pixel resolution, ensuring a uniform basis for evaluation across all samples.

\subsection{Image Preprocessing}

The preprocessing stage is critical for enhancing the features of MRI images and facilitating effective tumor detection. This process comprises several meticulously designed steps:

\begin{enumerate}
    \item \textbf{Smoothing with a Kernel:} The process begins with smoothing the image using a 7x7 kernel. This step reduces image noise and minor variations without significantly compromising important details. The kernel works by averaging the pixels within its area, effectively smoothing the image while preserving essential structures.
    
    \item \textbf{Bilateral Filtering:} Following smoothing, the image undergoes bilateral filtering. This advanced technique preserves edges while reducing noise. Unlike standard filters that only consider pixel proximity, bilateral filtering also accounts for intensity differences, ensuring that edges, such as those delineating tumors, remain sharp and distinct.
    
    \item \textbf{Grayscale Conversion:} The final step involves converting the image to grayscale. This simplification reduces the image's complexity by eliminating color variation, allowing the subsequent analysis to focus solely on intensity and contrast variations. Such focus is crucial for distinguishing between tumor and normal brain tissue, enhancing the model's ability to segment and identify tumors accurately.
\end{enumerate}

Each preprocessing step is designed to enhance the image to better highlight the features relevant to brain tumor classification, setting a strong foundation for the segmentation and deep learning stages that follow.

Smoothing: A 7x7 kernel is applied to smooth the image, reducing noise while preserving essential details.
Bilateral Filtering: To improve edge preservation, bilateral filtering is applied, enhancing the distinction between the tumor and surrounding tissues.
Grayscale Conversion: The processed image is converted to grayscale, simplifying the subsequent analysis by focusing on intensity variations rather than color.
\subsection{Segmentation using K-means Clustering}
Following preprocessing, K-means clustering with three clusters is employed to segment the image into distinct regions. This step is crucial for isolating tumor regions from the rest of the brain tissue, facilitating accurate classification by highlighting the tumor's boundaries and characteristics.

\subsubsection{Determining Optimal Clusters with the Elbow Method}
The Elbow Method was utilized to determine the optimal number of clusters for K-means clustering to ensure the most effective segmentation of brain MRI images. This method involves plotting the sum of squared distances for clusters and identifying the "elbow point," where the rate of decrease sharply changes. This point signifies the most appropriate number of clusters to use, balancing within-cluster variance against the total number of clusters. By applying the Elbow Method\cite{elbow}, we aimed to enhance the K-means algorithm's accuracy in delineating tumor boundaries, ensuring that the subsequent classification stages are founded on precise segmentation.

The within-cluster sum of squares (WCSS) is calculated using the formula:
\begin{equation}
    WCSS = \sum_{i=1}^{n}(x_i - c_i)^2
\end{equation}
where \( x_i \) represents a data point, \( c_i \) is the centroid of the cluster, and \( n \) is the number of data points in the cluster.

\begin{figure}[htbp]
\centerline{\includegraphics[width=0.5\textwidth]{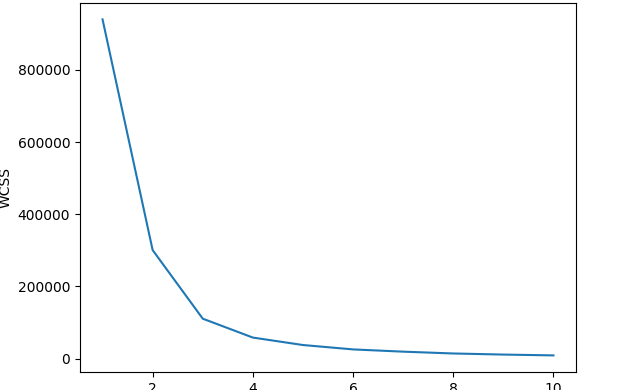}}
\caption{Elbow method result }
\label{fig}
\end{figure}

\subsection{Data Augmentation}
Data augmentation techniques are applied using the ImageDataGenerator class to counteract the limited size of the dataset and increase the model's generalization ability. Operations such as rescaling, shear transformation, zooming, and horizontal flipping are performed. Additionally, the preprocessing function myFunc is integrated into the data augmentation pipeline, ensuring that all images fed into the model are uniformly preprocessed.

\subsection{Model Architecture and Training}

The core of our classification framework is built upon the ResNet50 architecture, a pioneering model in deep learning that introduces the concept of residual learning to enable the training of much deeper networks. The architecture comprises:

\begin{itemize}
    \item \textbf{ResNet50 Base:} At its foundation, the model incorporates 48 Convolutional layers, along with 1 MaxPooling and 1 AveragePooling layer. Each Convolutional layer is equipped with batch normalization and ReLU activation, facilitating faster convergence and reducing the internal covariate shift. Shortcut connections are a hallmark of this architecture, addressing the degradation problem by enabling the unimpeded flow of gradients, which is crucial for training deep networks.
    
    \item \textbf{Flatten Layer:} Following the ResNet50 base, a Flatten layer transforms the multidimensional feature maps into a flat vector, bridging the fully connected layers.
    
    \item \textbf{Dense Layers:} The architecture concludes with two Dense layers. The initial Dense layer uses 128 units with ReLU activation, introducing non-linearity and the capacity to learn complex patterns. The final layer, consisting of a single unit with a sigmoid activation function, is dedicated to binary classification, outputting a probability indicating the presence of a tumor.
\end{itemize}

Compiled with the Adam optimizer and binary cross-entropy loss, the model is meticulously trained over 50 epochs. This structured approach leverages the depth and efficiency of ResNet50 and fine-tunes the network for the specific task of brain tumor classification from MRI images.

\subsection{Evaluation and Visualization}
The model's performance is evaluated upon training completion by visualizing accuracy and loss over epochs. These plots provide insights into the learning progression, indicating the model's ability to distinguish between normal and tumorous MRI images.

\section{Result}
\subsection{Overlap Analysis in Figshare Dataset}
In analyzing the overlap between the clustered regions and the actual tumor masks within the Figshare dataset, our study conducted a detailed comparison across the three tumor classes: Meningioma, Glioma, and Pituitary tumor. The comparison aimed to assess the effectiveness of K-means clustering in accurately identifying tumor regions. The Meningioma class showcased the highest unity, with the areas clustered closely matching the mask regions in almost all images. However, for the Glioma and Pituitary tumor classes, the analysis revealed several instances where the K-means clustering failed to detect the tumor, indicating a disparity in the algorithm's performance across different tumor types. This variability underscores the necessity for further refinement of clustering algorithms or exploring alternative methods for more consistent tumor detection across various classes.

The Intersection over Union (IoU) quantifies the overlap between the clustered regions and the tumor masks. The IoU is calculated using the formula\cite{iou}:

\begin{equation}
    IoU = \frac{|A \cap B|}{|A \cup B|}
\end{equation}

where \( A \) represents the predicted mask region and \( B \) represents the true mask region.

\begin{figure}[htbp]
\centerline{\includegraphics[width=0.5\textwidth]{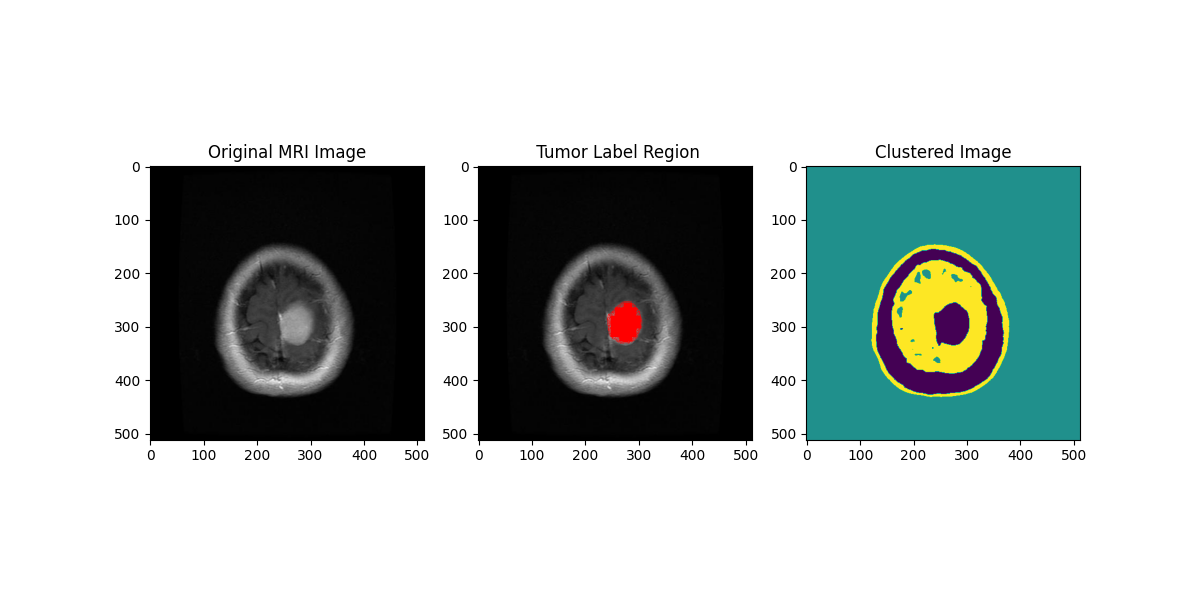}}
\caption{Comparision of Meningioma class }
\label{fig}
\end{figure}

\begin{figure}[htbp]
\centerline{\includegraphics[width=0.5\textwidth]{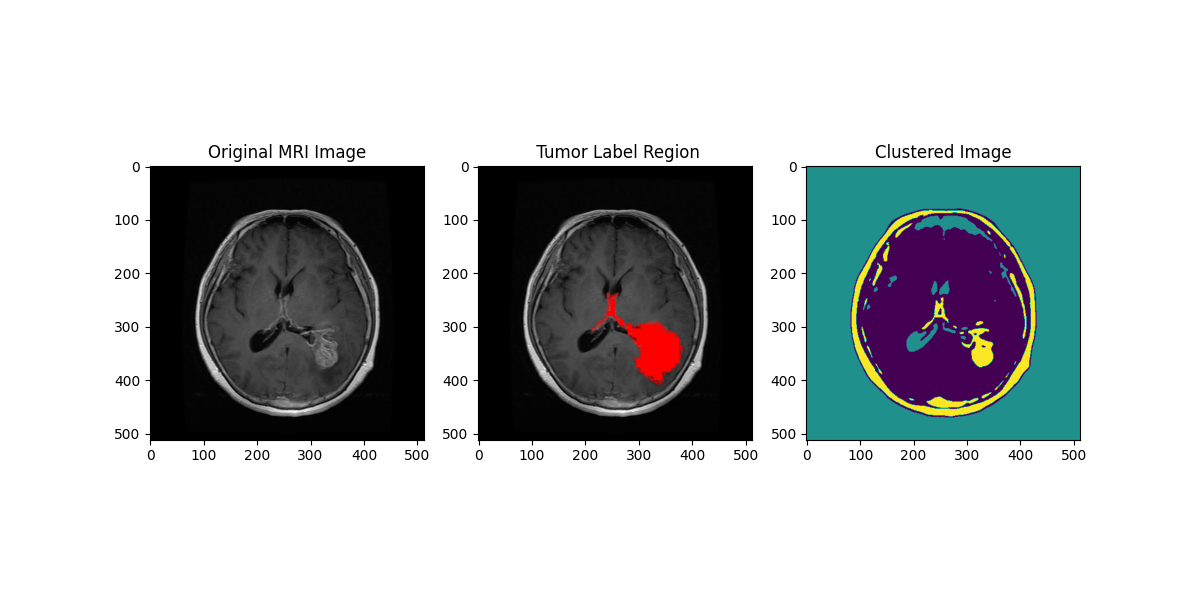}}
\caption{Glioma class showing 65\% overlap in tumor identification.}
\label{fig2}
\end{figure}

\begin{figure}[htbp]
\centerline{\includegraphics[width=0.5\textwidth]{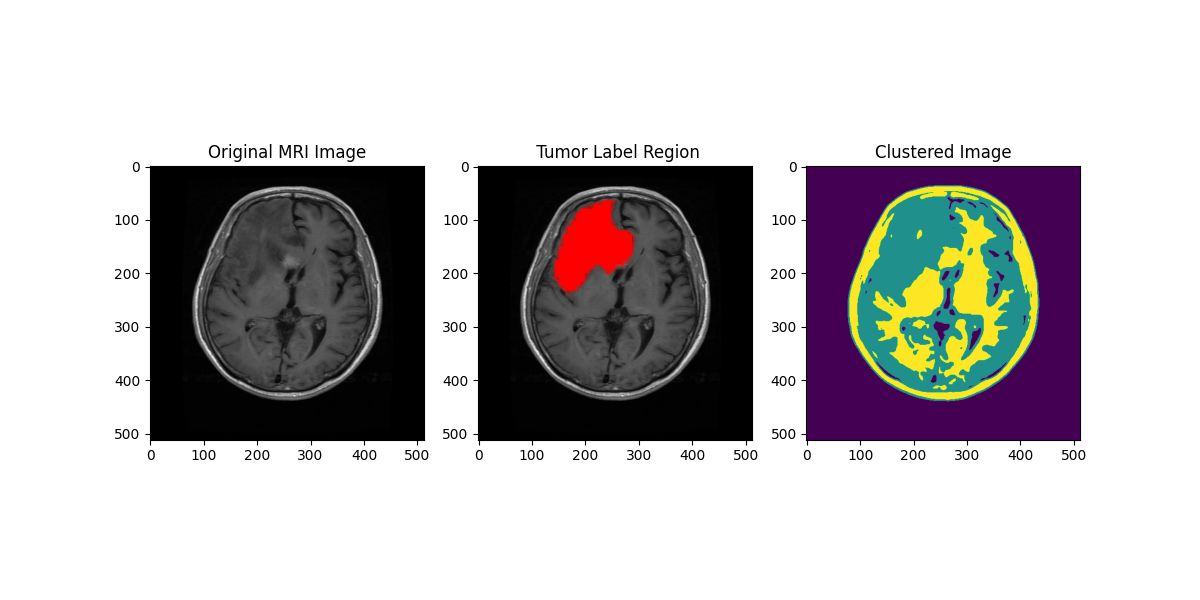}}
\caption{No Tumor detection on this image of the Glioma class.}
\label{fig3}
\end{figure}

\begin{figure}[htbp]
\centerline{\includegraphics[width=0.5\textwidth]{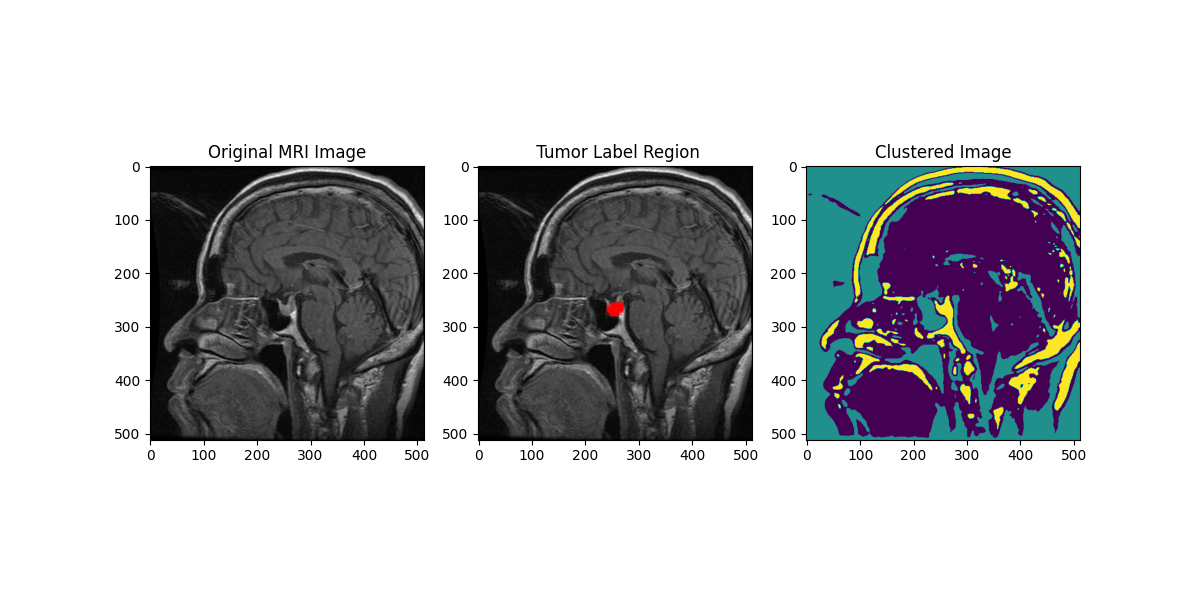}}
\caption{No Tumor detection on this image of the Pituitary class.}
\label{fig4}
\end{figure}

\subsection{Accuracy of Model Performance}
The model's performance on the Figshare dataset displayed progressive learning, as evidenced by the ascending training accuracy over the epochs. The Model Accuracy graph indicates a peak accuracy of 75\%, affirming the model's capability to identify tumor presence through MRI image classification. Conversely, test accuracy showed considerable variability, suggesting room for improvement in model generalization.

The Model Loss graph revealed a decrement in loss for both training and test sets, signifying the model's refinement in error minimization. Despite a notable spike in test loss mid-training, the general downward trend indicates a successful learning trajectory.

\begin{figure}[htbp]
\centerline{\includegraphics[width=0.5\textwidth]{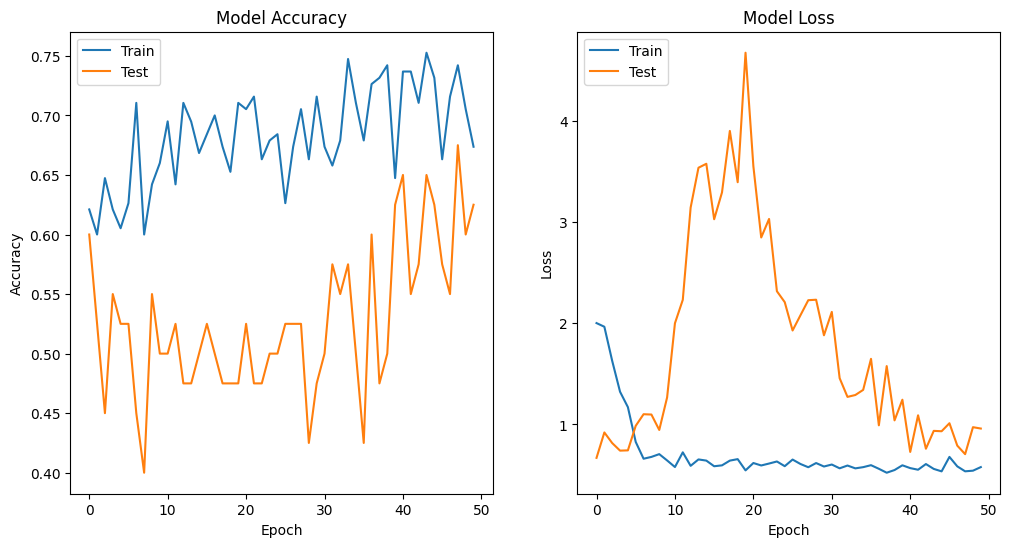}}
\caption{Accuracy of Model Performance}
\label{fig5}
\end{figure}

\section{Discussion}
This section presents a comparative analysis between the methodologies developed in this study and those reviewed in the Literature Review section. The following table encapsulates a comprehensive comparison, wherein each method, including our proposed approach, has been rigorously tested and benchmarked on the same Figshare dataset comprising 3064 images. This ensures a consistent and fair evaluation platform for all techniques under consideration, allowing for objectively assessing their relative performance in brain tumor classification tasks.

\begin{table}[h!]
\centering
\caption{Comparison of the proposed framework with the other state of art models}
\begin{tabular}{|p{0.3\linewidth}|p{0.3\linewidth}|p{0.2\linewidth}|}
\hline
\textbf{Method}  & \textbf{Classifier}  & \textbf{Accuracy} \\ \hline
Khan et al.\cite{brain1}  & SVM  & 97.8 \\ \hline
Soumik et al.\cite{brain2} & InceptionV3 & 99 \\ \hline
Irmak \cite{brain3}  & 3 different CNN & 99.33, 92.66, 98.14 \\ \hline
Dheepak et al.\cite{brain4}  & SVM  & 93 \\ \hline
Montoya et al.\cite{brain5} & Shallow neural networks (MLP)  & 86.82 \\ \hline
Proposed Method  & K-Means + ResNet50  & 75 \\ \hline
\end{tabular}
\label{table:comparison}
\end{table}

\subsection{Limitations and Future Work}

The current study has brought to light significant limitations in applying clustering techniques for MRI brain tumor detection. Though useful in differentiating between tissue types, the K-means clustering has not fully addressed the intricacies involved in tumor detection. This shortfall is reflected in the data presented in Figures 3, 4, and 5 in the Results section, where instances of non-detection of tumors through clustering have resulted in the introduction of inaccuracies into the subsequent training phase of the ResNet50 model. Such inaccuracies have manifested as elevated loss metrics and diminished accuracy rates, suggesting a discordance between the preprocessed data and the model's predictive capabilities.

In response to these challenges, future research directions will focus on incorporating advanced Edge Detection techniques\cite{edge}. These techniques aim to enhance the precision of tumor localization by delineating the perimeters more clearly than the current clustering methods allow. By integrating more accurate tumor boundary detection into the preprocessing phase, it is anticipated that the data fed into the classifier will be of higher fidelity, thereby reducing error rates and improving the classifier's performance. This approach is expected to refine the accuracy of tumor detection and contribute to developing more reliable automated diagnostic tools in the medical imaging field.

\section{Conclusion}

The exploration of deep learning models in this study, particularly the integration of preprocessing methods with Convolutional Neural Networks (CNNs), has revealed strengths and areas for improvement in medical imaging for tumor detection. The successful application of models such as SVM and InceptionV3 has set a high benchmark in accuracy. Yet, the limitations in clustering techniques, as demonstrated by our model's lower accuracy, suggest that more nuanced approaches are necessary for complex tasks like brain tumor detection.

While the proposed model has shown promise, its performance is hindered by the inherent limitations of K-means clustering in effectively segmenting tumor regions, as evidenced by the results. This shortfall emphasizes the need for more sophisticated segmentation methods to capture the subtle nuances of medical images more effectively.

Future research will focus on overcoming these limitations by exploring advanced preprocessing techniques, such as Edge Detection, which could provide more precise tumor boundary delineation. Improving the feature extraction phase is expected to lower loss rates and enhance the overall performance of the CNN model. It's also imperative to consider other neural network architectures and learning paradigms that may offer better generalization and robustness for medical image analysis.

\end{document}